\documentclass[pra,twocolumn,superscriptaddress,10pt,longbibliography]{revtex4-1}

\usepackage[utf8]{inputenc}
\usepackage[T1]{fontenc}
\usepackage{amsmath}    
\usepackage{graphicx}   
\usepackage{verbatim}   
\usepackage{color}      
\usepackage{subfigure}  
\usepackage{braket}
\usepackage[hidelinks]{hyperref}
\usepackage[capitalize]{cleveref}
\usepackage{booktabs}
\usepackage{siunitx}
\DeclareSIUnit\gauss{G}
\DeclareSIUnit\bohr{a_{B}}
\usepackage{bbm}
\usepackage{ulem}

\usepackage{color}
\definecolor{mygreen}{rgb}{0,0.5,0}
\definecolor{mygrey}{rgb}{0.5,0.5,0.5}
\definecolor{myred}{rgb}{0.75,0,0}
\definecolor{myblue}{rgb}{0,0,0.75}
\definecolor{mymagenta}{cmyk}{0,1,0,0.12}
\definecolor{mycyan}{cmyk}{1,0,0,0.12}
\definecolor{myorange}{rgb}{1,0.5,0}
\definecolor{myviolet}{rgb}{0.5,0.0,0.75}

\newcommand{\ICFOAddress}{ICFO-Institut de Ci\`encies Fot\`oniques, The Barcelona Institute of Science and Technology, 08860 Castelldefels (Barcelona), Spain} 
\newcommand{\ICREAAddress}{ICREA -- Instituci\'{o} Catalana de Recerca i Estudis Avan\c{c}ats, 08010 Barcelona, Spain}
\newcommand{\QusideAddress}{Quside Technologies S.L., C/Esteve Terradas 1, Of. 217, 08860 Castelldefels (Barcelona), Spain}

\begin{document}

\title{Interferometric measurement of interhyperfine scattering lengths in $^{87}$Rb}
\author{Pau Gomez}
\affiliation{\ICFOAddress}

\author{Chiara Mazzinghi}
\affiliation{\ICFOAddress}

\author{Ferran Martin}
\affiliation{\ICFOAddress}
\affiliation{\QusideAddress}

\author{Simon Coop}
\affiliation{\ICFOAddress}

\author{Silvana Palacios}
\affiliation{\ICFOAddress}

\author{Morgan W. Mitchell}
\affiliation{\ICFOAddress}
\affiliation{\ICREAAddress}

\date{\today}

\begin{abstract}
We present interferometeric measurements of the $f=1$ to $f=2$ inter-hyperfine scattering lengths in a single-domain spinor Bose-Einstein condensate of $^{87}$Rb. The inter-hyperfine interaction leads to a strong and state-dependent modification of the spin-mixing dynamics with respect to a non-interacting description. We employ hyperfine-specific Faraday-rotation probing to reveal the evolution of the transverse magnetization in each hyperfine manifold for different state preparations, and a comagnetometer strategy to cancel laboratory magnetic noise. The method allows precise determination of inter-hyperfine scattering length differences, calibrated to intra-hyperfine scattering length  differences. We report $(a_{3}^{(12)}-a_{2}^{(12)})/(a_{2}^{(1)}-a_{0}^{(1)})=\SI{-1.27(15)}{}$ and $(a_{1}^{(12)}-a_{2}^{(12)})/(a_{2}^{(1)}-a_{0}^{(1)})=\SI{-1.31(13)}{}$, limited by atom number uncertainty. With achievable  control of atom number, we estimate precisions of $ \SI{\approx 0.3}{\percent}$ should be possible with this technique.  

\end{abstract}

\maketitle

\section{Introduction}

Since the advent of Bose-Einstein condensation (BEC) in ultracold quantum gases, experimental access to the spin degrees of freedom and resulting spin-dependent interactions have expanded greatly. The pioneering $^{87}$Rb, $^{23}$Na and $^7$Li experiments \cite{95Cornell,95Ketterle,95Bradley} used magnetic trapping that restricted their studies to scalar BECs in {low field seeking} Zeeman sublevels. By introducing optical trapping techniques \cite{98Ketterle,01Chapman}, the spin degree of freedom became accessible. This enabled the study of spin-mixing dynamics \cite{04Chapman,04Sengstock,05Chapman,12Dalibard}, spontaneous magnetic symmetry breaking \cite{06Stamper-Kurn,10Stamper-Kurn,13Arlt}, domain formation \cite{98Ketterle,06Stamper-Kurn,14Spielman} and exotic topological spin excitations \cite{12Yong-il,14Hall} in spinor Bose-Einstein condensates (SBEC). 

These rich dynamics arise from the interplay between superfluidity and magnetism, which for a single, spin-$f$ species and s-wave binary contact interactions are described by $f+1$ parameters, the \textit{intrahyperfine} scattering lengths.  In the case  of $^{87}$Rb, these have been separately determined for the $f=1$ and $f=2$ ground-state manifolds \cite{02Verhaar, 05Chapman, 06Bloch}.   \textit{Interhyperfine} interactions are less well studied, but nonetheless play an important role in determining the miscibility of multiple BEC species \cite{08Wieman, 08Inguscio, 11Cornish}, and have been used to produce spin-squeezing with its attendant entanglement, and Bell-type correlations \cite{14Oberthaler,15Oberthaler,16Treutlein, 18Treutlein,18Klempt}. For $^{87}$Rb, the full set of inter-hyperfine spin interaction parameters has recently been measured \cite{18Hirano} with intriguing results. The current best values indicate that in an equal $f=1$, $f=2$ ground-state mixture, the $f=1$ component manifests a polar ground state at zero magnetic field \cite{18Saito} even though the  $f=1$ component alone is ferromagnetic \cite{04Chapman}.  

In this work we report precision measurements on the $^{87}$Rb inter-hyperfine $f=1\leftrightarrow f=2$ scattering lengths, using a novel comagnetometer strategy. We use a single-domain SBEC \cite{18Palacios}, with non-destructive Faraday probing \cite{10Mitchell} for simultaneous readout of amplitude and phase of the transverse magnetization in $f=1$ and $f=2$. The method is interferometric: the scattering of interest induces a phase shift among the Zeeman levels, which is detected via the precession angle. The observed dynamics are compared to mean-field simulations under the single-mode approximation (SMA) \cite{99Bigelow, 12Ueda}, yielding the two spin-dependent inter-hypefine interaction parameters \cite{18Saito}.

The presentation is organized as follows: Section \ref{sec:hamiltonian} describes the interhyperfine interaction  for $^{87}$Rb. It discusses the simplifications under the rotating wave approximation (\cref{sec:RWA}) and the implementation of the numerical simulations (\cref{sec:NumericalIntegration}). Data interpretation and error sources are detailed in \cref{sec:errEstimate}. \Cref{sec:setup} and \cref{sec:calib} introduce the experimental setup and required classical calibrations. Section \ref{sec:inter-hyperfine} describes the measurement of the spin-dependent interaction parameters. In \cref{sec:results} we present the resulting inter-hyperfine scattering lengths and compare against literature values.  

\section{Mean-field description}\label{sec:hamiltonian}
A SBEC can be described by a vectorial order parameter, which in the SMA can be written
\begin{equation}\label{eq:sma}
\Psi_m^{(f)}(\mathbf{r},t)=\Psi_{\rm SMA}(\mathbf{r})\cdot\xi_m^{(f)}(t)\; ,
\end{equation}
where $f=1,2$ and $m=-f,...,f$. The spin-independent spatial wave function $\Psi_{\rm SMA}(\mathbf{r})$ and the relative spin amplitudes $\xi_m^{(f)}$ are normalized as follows:
\begin{subequations} \label{eq:normalization}
\begin{align}
\label{eq:normalizationSpatial}
\int d^3r |\Psi_{\rm SMA}(\mathbf{r})|^2 &= 1\; \\ 
\sum_{f,m}|\xi_m^{(f)}|^2 &=N\; ,
\end{align}
\end{subequations}
where $N$ is the number of atoms. For BECs significantly larger than the density healing length, the kinetic contribution to the total energy is negligible and the density distribution is described by a Thomas-Fermi profile \cite{96Pethick,96Stringari,97Smith}:
\begin{equation}\label{eq:TFApprox}
N|\Psi_{\rm SMA}(\mathbf{r})|^2 =
\left\{ 
\begin{array}{lr}
\frac{\mu-V(\mathbf{r})}{g_0^{(1)}}  & \text{when}\;   V({\bf r}) < \mu \\
0 &{\rm otherwise}
\end{array}
\right.
\end{equation}
where $V(\mathbf{r})$ is the underlying trapping potential and  $g_0^{(1)}$ the spin-independent interaction coefficient for $f=1$ (see eq. (\ref{eq:scattCoefg01}) below). The chemical potential $\mu$ is obtained by normalizing the spatial wave function as defined in \cref{eq:normalizationSpatial}. 

In the SMA, the spatial dependence of the wave function is \textit{integrated out} and only contributes through the effective volume $V_{\rm eff}=(\int dr^3 |\Psi_{\rm SMA}(\mathbf{r})|^4)^{-1}$. For the density profile in \cref{eq:TFApprox} and a harmonic trapping potential with mean trapping frequency $\bar{\omega}$, the effective trapping volume becomes: 
\begin{align}
\label{eq:Veff}
V_{\rm eff}&=\frac{14}{15}\pi \bar{R}_{\rm TF}^3\; ,\\[5pt]
\label{eq:RTF}
\bar{R}_{\rm TF}&=\left(\frac{15}{4\pi}\frac{g_0^{(1)}N}{M \bar{\omega}^2}\right)^{1/5}\; ,
\end{align}
where $\bar{R}_{\rm TF}$ is the mean Thomas-Fermi radius and $M$ the atomic mass. 

\newcommand{\ftot}{f_{\rm tot}}
\renewcommand{\ftot}{c}

We follow the notation in \cite{18Saito} and write s-wave scattering lengths as $a_{\ftot}^{({\cal F})}$.  The scattering channel $\ftot$ specifies the total spin quantum number of the colliding atoms, while ${\cal F} \in\lbrace1,2,12\rbrace$ indicates $f=1$ or $f=2$ intrahyperfine or  $f=1,2$ interhyperfine scattering, respectively. In terms of these are defined the interaction coefficients that appear in the single-mode description:
\begin{subequations}
\begin{align}
\label{eq:scattCoefg01}
g_{0}^{(1)}  =& \frac{4\pi\hbar^2}{M}\frac{a_0^{(1)}+2a_2^{(1)}}{3} \\
g_1^{(1)}  =& \frac{4\pi\hbar^2}{M}\frac{a_{2}^{(1)}-a_{0}^{(1)}}{3}\; ,\\
g_1^{(2)}  =& \frac{4\pi\hbar^2}{M}\frac{a_{4}^{(2)}-a_{2}^{(2)}}{7}\; ,\\
g_2^{(2)}  =& \frac{4\pi\hbar^2}{M}\frac{7a_{0}^{(2)} -10 a_{2}^{(2)}+3a_{4}^{(2)}}{7}\; ,\\
\label{eq:scattCoefg012}
g_0^{(12)}  =& \frac{4\pi\hbar^2}{M}\frac{2a_2^{(12)} + a_3^{(12)}}{3} ,\\
\label{eq:scattCoefg112}
g_1^{(12)}  =& \frac{4\pi\hbar^2}{M}\frac{a_{3}^{(12)}-a_{2}^{(12)}}{3}\; ,\\
\label{eq:scattCoefg212}
g_2^{(12)}  =& \frac{4\pi\hbar^2}{M}\frac{3a_{1}^{(12)}-5a_{2}^{(12)}+2a_3^{(12)}}{3}\; .
\end{align}
\end{subequations}

The $f=1$ manifold contributes an energy 
\begin{equation}\label{eq:E1}
E^{(1)}=\sum_{m}(p^{(1)}m+ q^{(1)}m^2)  \left\lvert\xi_m^{(1)} \right \lvert^2+\frac{1}{2V_{\rm eff}}g_1^{(1)}\mathbf{F}^{(1)}\cdot \mathbf{F}^{(1)} \; ,
\end{equation}
where $p^{(1)}$ and $q^{(1)}$ describe the linear and quadratic  Zeeman shifts (LZS and QZS, respectively), and 
 $\mathbf{F}^{(f)}$ is the mean spin vector with cartesian components $F_i^{(f)} \equiv \mathbf{\xi}^{(f) \dagger} \hat{F}_i^{(f)} \mathbf{\xi}^{(f)}$, where $\hat{F}_i^{(f)}$ are spin-$f$ matrices. 
\\
 
The $f=2$ manifold contributes an energy 
\begin{align}\label{eq:E2}
E^{(2)}=& \sum_{m}(p^{(2)}m+ q^{(2)}m^2)  \left\lvert\xi_m^{(2)} \right \lvert^2  \nonumber \\
&+\frac{1}{2V_{\rm eff}} \left(  g_1^{(2)}\mathbf{F}^{(2)}\cdot \mathbf{F}^{(2)} +g_2^{(2)}\left\lvert A_0^{(2)} \right \lvert^2\right)\; ,
\end{align}
where $p^{(2)}$ and $q^{(2)}$ describe the LZS and QZS of the $f=2$ manifold, and $A_0^{(2)}$ is the spin-singlet scalar 
\begin{equation}
A_0^{(2)} \equiv \frac{1}{\sqrt{5}}\left( 2\xi_2^{(2)}\xi_{-2}^{(2)} - 2\xi_1^{(2)}\xi_{-1}^{(2)} +\xi_0^{(2)}\xi_{0}^{(2)}\right).
\end{equation}

The inter-hyperfine scattering contribution has been recently described \cite{18Saito} and can be written 
\begin{align}\label{eq:E12noaverage}
E^{(12)}= &\frac{1}{V_{\rm eff}} \left(g_0^{(12)} \sum_{m'} |\xi_{m'}^{(1)}|^2  \sum_{m''} |\xi_{m''}^{(2)}|^2 \right. \nonumber\\
 &\left.+  g_1^{(12)} {\bf F}^{(1)}\cdot {\bf F}^{(2)}+ g_2^{(12)}P_1^{(12)}\right),
\end{align}
where 
\begin{align}\label{eq:P1noaverage}
P_1^{(12)} & =  \left| \sqrt{\frac{1}{10}} \xi_1^{(1)} \xi_0^{(2)} -   \sqrt{\frac{3}{10}}  \xi_0^{(1)} \xi_1^{(2)}   +   \sqrt{\frac{3}{5}} \xi_{-1}^{(1)} \xi_2^{(2)} \right|^2
\nonumber \\ & + 
\left| \sqrt{ \frac{3}{10}} \xi_1^{(1)} \xi_{-1}^{(2)} -\sqrt{  \frac{2}{5}} \xi_0^{(1)} \xi_0^{(2)} + \sqrt{   \frac{3}{10}} \xi_{-1}^{(1)} \xi_1^{(2)} \right|^2
 \nonumber \\ & + 
\left \vert\sqrt{  \frac{3}{5}} \xi_1^{(1)} \xi_{-2}^{(2)} -\sqrt{   \frac{3}{10}}  \xi_0^{(1)} \xi_{-1}^{(2)} + \sqrt{   \frac{1}{10}} \xi_{-1}^{(1)} \xi_0^{(2)} \right|^2 \hspace{3mm}
\end{align}
results from inter-hyperfine scattering with total quantum number $c=1$. 

\newcommand{\pmin}{p_0}
\newcommand{\poff}{p_s}
\newcommand{\supone}{^{(1)}}
\newcommand{\suptwo}{^{(2)}}
\newcommand{\suponetwo}{^{(12)}}
\newcommand{\supf}{^{(f)}}

\subsection{Rotating wave approximation}\label{sec:RWA}
The LZS terms $p^{(1)}$ and $p^{(2)}$ induce Larmor precession of the spins about the magnetic field direction, assumed to be ${\bf z}$, the same as the quantization axis. ${\bf F}\supone$ and ${\bf F}\suptwo$ precess in opposite senses and with nearly equal angular frequency:  $p\supone = -\pmin - \poff$, $p\suptwo = \pmin - \poff$, where in $^{87}$Rb  $\pmin/(Bh) \approx  \SI{700}{\kilo\hertz\per\gauss}$ and $\poff/(Bh) \approx \SI{1.39}{\kilo\hertz\per\gauss}$. It is natural to work in a dual-rotating frame defined by $\xi\supone_m \rightarrow \xi\supone_m \exp[i m \pmin/\hbar]$, $\xi\suptwo_m \rightarrow \xi\suptwo_m \exp[-i m \pmin t/\hbar]$, with the consequence $p\supone \rightarrow  - \poff$, $p\suptwo \rightarrow  - \poff$. We note that rotation-invariant terms such as ${\bf F}\supone \cdot {\bf F}\suptwo$ are unaffected by this change of frame.  In contrast, many inter-hyperfine interaction terms acquire an oscillating factor, e.g. $\xi\supone_{-1} \xi\suptwo_{1} \rightarrow \xi\supone_{-1} \xi\suptwo_{1}\exp[ i 2\pmin t/\hbar]  $ .  

In the experiments described below, the precession frequency $\pmin/h \sim \SI{100}{\kilo\hertz}$ is much faster than the collisional spin dynamics, e.g.  $|Ng_{1}^{(1)}/(V_{\rm eff}h)| \sim \SI{3}{\hertz}$. This motivates the rotating wave approximation (RWA), i.e. dropping the rapidly oscillating terms. From perturbation theory we expect the RWA to introduce a fractional error at the $10^{-4}$ level, which is negligible in this context.   

Under this simplification, and excluding the constant term $\propto g_0^{(12)}$, the inter-hyperfine energy becomes:

\begin{equation}\label{eq:E12}
\overline{E^{(12)}}=\frac{1}{V_{\rm eff}} \left( g_1^{(12)} F_z^{(1)}F_z^{(2)}+ g_2^{(12)} \overline{P_1^{(12)}}\right)\; ,
\end{equation}
where 
\begin{align}\label{eq:P1}
 \overline{P_1^{(12)}} = & \frac{1}{10}\left \vert \xi_1^{(1)} \xi_0^{(2)} \right\vert^2   +   \frac{3}{10}\left \vert \xi_0^{(1)} \xi_1^{(2)} \right\vert^2  +   \frac{3}{5}\left \vert \xi_{-1}^{(1)} \xi_2^{(2)} \right\vert^2
\nonumber \\ & + 
 \frac{3}{10}\left \vert \xi_1^{(1)} \xi_{-1}^{(2)} \right\vert^2   +   \frac{2}{5}\left \vert \xi_0^{(1)} \xi_0^{(2)} \right\vert^2  +   \frac{3}{10}\left \vert \xi_{-1}^{(1)} \xi_1^{(2)} \right\vert^2
 \nonumber \\ & + 
  \frac{3}{5}\left \vert \xi_1^{(1)} \xi_{-2}^{(2)} \right\vert^2   +   \frac{3}{10}\left \vert \xi_0^{(1)} \xi_{-1}^{(2)} \right\vert^2  +   \frac{1}{10}\left \vert \xi_{-1}^{(1)} \xi_0^{(2)} \right\vert^2. \hspace{3mm}
\end{align}
\begin {table}[t!]
\begin{center}
 \begin{tabular}{p{2.0cm} p{2.5cm} p{2.3cm} p{1cm}} 
 \hline
 Parameter  &  Mean Value & Uncertainty  &  Ref \\ 
 \hline\hline 
 \addlinespace[0.5ex]
\addlinespace[0.5ex]
$N$ & see sec. \ref{sec:calib}, \ref{sec:inter-hyperfine}& \SI{10}{\%} &  \\
\addlinespace[0.5ex]
$a_0^{(1)}$ & \SI{101.8}{\bohr} & \SI{0.2}{\bohr} &\cite{02Verhaar}  \\
\addlinespace[0.5ex]
$a_{2}^{(1)}-a_{0}^{(1)}$ & \SI{-1.07}{\bohr} & \SI{0.09}{\bohr}&\cite{06Bloch}  \\ 
\addlinespace[0.5ex]
$a_{2}^{(2)}-a_{0}^{(2)}$ & \SI{3.51}{\bohr} & \SI{0.54}{\bohr}&\cite{06Bloch}  \\ 
\addlinespace[0.5ex]
$a_{4}^{(2)}-a_{2}^{(2)}$ & \SI{6.95}{\bohr} & \SI{0.35}{\bohr}&\cite{06Bloch}  \\ 
\addlinespace[0.5ex]
\hline
\end{tabular}
\end{center}
\caption{Mean values and associated uncertainties for the atom number (N) and intra-hyperfine scattering lengths in terms of the Bohr radius $\SI{}{\bohr}$. The 10\% standard deviation in the atom number is a conservative estimate of the measured atom number fluctuations within repetitions of the same experimental sequence. These arise from long-term atom number drifts and stochastic loading in the experimental sequences of \cref{sec:calib} and \cref{sec:inter-hyperfine}. } 
\label{tab:scattLen}
\end{table}

\subsection{Numerical integration}
\label{sec:NumericalIntegration}
Once the intra- and inter-hyperfine contributions have been obtained, the dynamical evolution of the spin amplitudes $\xi_m^{(f)}$ are computed by differentiating the total energy: 
\begin{equation}\label{eq:GPE}
i\hbar \frac{\partial \xi_m^{(f)}}{\partial t}=\frac{\delta (E)}{\;\delta \xi_m^{(f)\; * }}\; ,
\end{equation}
where $E=E^{(1)}+E^{(2)}+\overline{E^{(12)}}$. The right-hand side of \cref{eq:GPE} is computed analytically and numerical integration (via the ODEPACK routine LSODA) is used to solve the resulting set of eight coupled differential equations \cite{GPE_F1F2}. 

\subsection{Data interpretation and error estimates}
\label{sec:errEstimate}
In Sections~\ref{sec:calib}, \ref{sec:inter-hyperfine1} and \ref{sec:inter-hyperfine2} we fit the model dynamics of \cref{eq:GPE} to observed data, with the intent to calibrate $q^{(f)}$ and $\bar{\omega}_{\rm{eff}}$ (effective trapping frequency), and determine $g_1^{(12)}$ and  $g_2^{(12)}$, or equivalently $a_{3}^{(12)}-a_{2}^{(12)}$ and $a_{1}^{(2)}-a_{2}^{(2)}$.  The relative intra-hyperfine scattering lengths $a_{2}^{({1})}-a_{0}^{({1})}$, $a_{2}^{({2})}-a_{0}^{({2})}$ and $a_{4}^{({2})}-a_{2}^{({2})}$ also appear as parameters in \cref{eq:GPE}. Their literature values and associated uncertainties are shown in \cref{tab:scattLen}. Note that theory and experiment are at present discrepant for $a_{2}^{(1)}-a_{0}^{(1)}$ \cite{05Chapman,06Bloch,02Verhaar}, which introduces a systematic uncertainty into our fit results.

A numerical exploration of the dependence of the fitted values for $\bar{\omega}_{\rm{eff}}$,  $g_1^{(12)}$ and  $g_2^{(12)}$ finds that in each case only $a_{2}^{(1)}-a_{0}^{(1)}$  contributes an uncertainty that is significant on the scale of the experimental precision. For the inter-hyperfine interaction terms, that dependence is linear and we report ratios of the form $(a_{3}^{(12)}-a_{2}^{(12)})/(a_{2}^{(1)}-a_{0}^{(1)})$, where the numerator is the fit result and the denominator is a fixed parameter in \cref{eq:GPE}. Similarly, we report $(a_{1}^{(12)}-a_{2}^{(12)})/(a_{2}^{(1)}-a_{0}^{(1)})$, $g_1^{(12)}/g_1\supone$, and $g_2^{(12)}/g_1\supone$. These ratios, unlike the fit result itself, are insensitive to the value of $a_{2}^{(1)}-a_{0}^{(1)}$, again at the level of precision of the experimental results. The dependence of the fitted $\bar{\omega}_{\rm{eff}}$ on $a_{2}^{(1)}-a_{0}^{(1)}$ is described in \cref{sec:calib}.

While systematic errors in the atom number readout are calibrated in \cref{sec:calib}, a remaining uncertainty arises from experimental atom numbers fluctuations and drifts. Atom numbers and their fluctuations were estimated by repeated trap loading, state preparation, and destructive absorption imaging prior to acquiring data runs such as the one reported in \cref{fig:trapFreq}. Despite this, a significant uncertainty accrues due to drifts in the $^{87}$Rb background pressure.  We account for this with a systematic uncertainty of $\pm10\%$ rms deviation around the measured atom numbers. The value $\pm10\%$ describes the observed drifts from run to run, as well as the observed fluctuations of $f=2$ population shown in \cref{fig:scattLen1}.

For most quantities derived from this analysis, we report the statistical averages and standard deviations of the corresponding fit parameters. For the transverse magnetization $F_\perp^{(f)}$ we report the median and 90\% confidence interval, which is more meaningful as $F_\perp^{(f)}$ is intrinsically positive-valued and asymmetrically distributed.

\section{ Experimental setup}\label{sec:setup}
\begin{figure}[t!]
\includegraphics[width=1\linewidth]{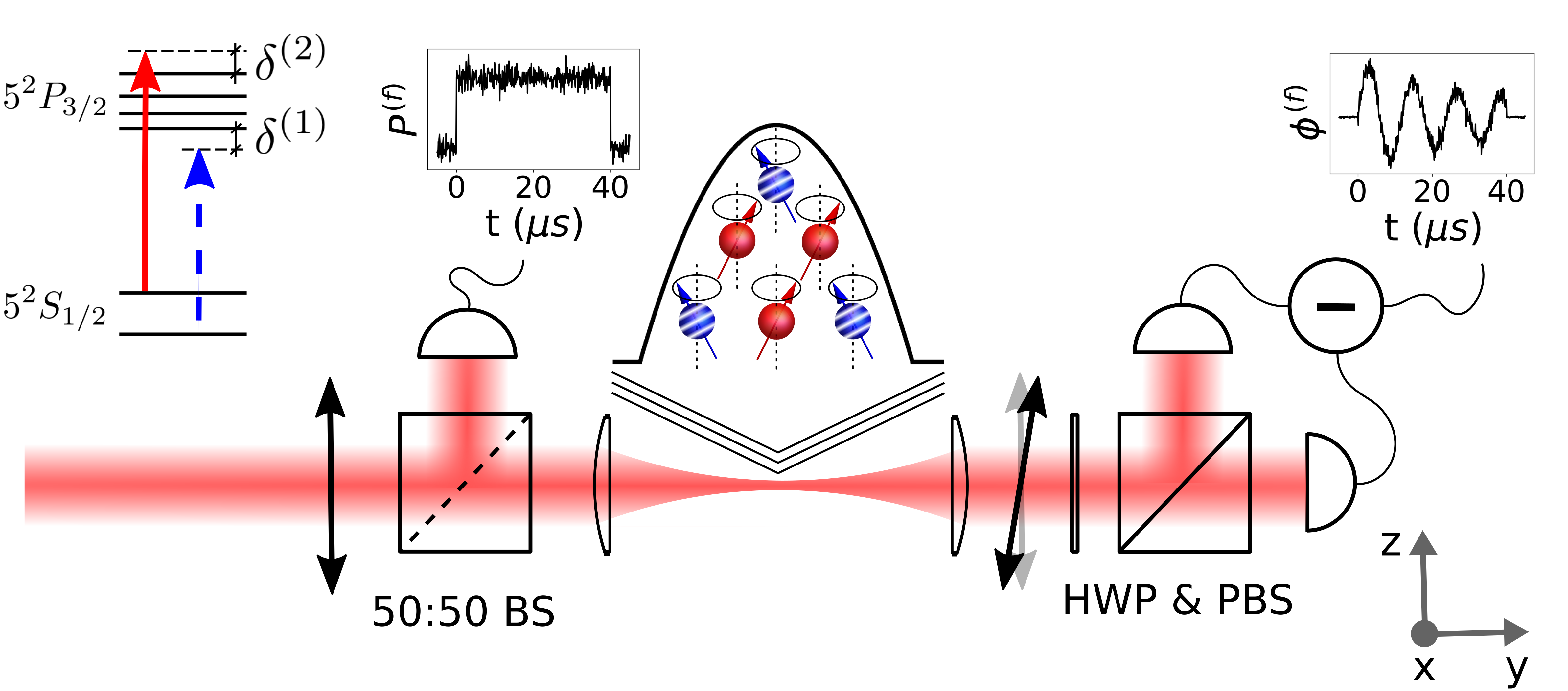} 
\caption{Faraday probing of a SBEC formed by $f=1$ (blue, striped) and $f=2$ (red, solid) atoms precessing around the applied magnetic field ${\bf B} = B {\bf z}$. Hyperfine selective probing is achieved by sending light closely detuned to $1\leftrightarrow 0'$ or $2\leftrightarrow 3'$ for $f=1$ and $f=2$, respectively. D2 line transitions, probe frequencies and detunings $\delta^{(1)}$ and $\delta^{(2)}$ are depicted on the left. The dispersive interaction rotates the probe beam by an angle proportional to the atomic spin projection $\phi^{(f)}=\frac{1}{2}G_1^{(f)}F_y^{(f)}$. The rotation angle is recorded on a differential photodetector, which has been balanced by adjusting the $\lambda/2$-waveplate (HWP) before the polarizing beam-splitter (PBS). The probe light power $P^{(f)}$ is monitored via a photodetector behind a 50:50 beam-splitter (BS) at the beginning of the optical path.  }	
\label{fig:setup}
\end{figure}

The experiments have been performed in a SBEC of $^{87}$Rb. After \SI{4.5}{\second} of all-optical evaporation in a crossed-beam optical dipole trap, a SBEC with typically $10^5$ atoms is achieved. Due to spatial bunching, thermal-thermal collisions are expected to contribute twice the per-atom energy of condensate-condensate or thermal-condensate collisions \cite{KetterlePRA1997, HarberPRA2002}. Although no thermal fraction is observed, we estimate that for a conservative upper bound of 10\% thermal fraction, the additional contribution of the thermal cloud is a $\SI{\sim 1}{\%}$ effect on the measured intra- and intrer-hyperfyne dynamics, which will henceforth be neglected. During the spin-dynamics phase of the experiment, the trap conditions satisfy both static  \cite{Ho1998,Law1998SMA}  and dynamic \cite{MakelaPRA2011, MakelPRA2012} criteria for stability of a single spin domain, and long-time spin relaxation measurements confirm the expected single-domain behavior \cite{18Palacios}.

The spin state of the atoms can be probed by Stern-Gerlach imaging or non-destructive Faraday probing, shown in \cref{fig:setup}. In the later case, a probe beam focused to a few times the Thomas-Fermi radius separately probes the transverse magnetization in $f=1$ and $f=2$. By alternating between light closely detuned to $1\leftrightarrow 0'$ or $2\leftrightarrow 3'$ ($D_2$ line transitions), we either interrogate the $f=1$ or the $f=2$ manifold. The probing pulses are linearly polarized and experience a rotation $\phi^{(f)}$, which is proportional to the atomic spin projection along their propagation direction ($\mathbf{y}$ axis). Under the influence of externally applied magnetic fields (along $\mathbf{z}$ axis), the LZS terms in \cref{eq:E1} and \cref{eq:E2} induce rapid Larmor precessions of the transverse spin and the Faraday rotation signal is of the form \cite{18Palacios}:  
\begin{align}\label{eq:faradaySignal}
\phi^{(f)}[t] =&\frac{1}{2}G_1^{(f)}F_y^{(f)}[t]\;, \\
F_y^{(f)}[t] =&F_{\perp}^{(f)} e^{-t/t_{\rm{dep}}^{(f)}}  \sin\left[ \frac{p^{(f)}}{\hbar} t + \theta^{(f)} \right],\label{eq:faradaySignal2}
\end{align}
where time $t$ is referenced to the start of the Faraday probing pulse. The vector atom-light coupling factor $G_1^{(f)}$ depends on the detuning $\delta^{(f)}$ to the above mentioned transitions and will be specified for the different experimental sequencs of this work.  The polarization rotation is continuously monitored for several Larmor periods and recorded on a balanced differential photodetector \cite{16Mitchell}. The amplitude and initial phase of the obtained oscillatory traces reveal the amplitude $F_\perp^{(f)}$ and precession angle $\theta^{(f)}$ of the transverse magnetization in $f=1$ and $f=2$. Depolarization caused by off-resonant photon absorption events results in a characteristic depolarization time $t_{\rm{dep}}^{(f)}$ \cite{EchanizJO2005, KubasikPRA2009}.
\\
The above-described Faraday probing setup is operated in the photon shot-noise limited regime \cite{16Mitchell}. In this regime, the readout noise of the transverse spin components, $\Delta F_x^{(f)}$ and $\Delta F_y^{(f)}$ is
\begin{equation}
\Delta F_x^{(f)}=\Delta F_y^{(f)}=\frac{1}{| G_1^{(f)}|}\sqrt{\frac{2}{N_L^{(f)}}}\;,
\end{equation}
where $N_L^{(f)}$ is total number of photons for Faraday probing the hyperfine manifold $f$. For the photon numbers and atom-light coupling factors of this work, the readout noise is estimated to $\SI{\approx 1000}{}$ spins.

\section{Calibration of trap conditions}\label{sec:calib}

For a precise determination of the inter-hyperfine scattering parameters, we require best-estimate values and uncertainties for the experimental parameters that appear in \cref{sec:hamiltonian}. These are the QZS $q^{(f)}$, the mean trapping frequency $\bar{\omega}$ and the atom number $N$. Precise knowledge of the LZS is not required, because the signals are either insensitive to the Larmor precession angles $\theta^{(1)}$ and  $\theta^{(2)}$, or are sensitive only to their sum, to which the net LZS contribution is small. The LZS must, however, be large enough that the RWA is valid. 

There are multiple sources for systematic uncertainties in the above mentioned parameters. The QZS is potentially affected by tensorial light shifts caused by the intense trapping beams \cite{17Coop}. The trapping frequency depends on power levels and precise alignment of the crossed dipole traps, and is tipically calibrated \textit{in situ}. The inferred atom number is sensitive to the magnification and polarization of the absorption imaging light, as well as to  the absorption cross section. For an absolute calibration of the measured atom numbers, schemes based on projection noise scaling in SBECs have been reported \cite{Koschorreck2010, 13Muessel}. 

We note that $q^{(f)}$, $\bar{\omega}$ and $N$ enter into $f=1$ spin dynamics and $f=1,2$ inter-hyperfine spin dynamics in the same way, which provides an opportunity to calibrate the net effect of these variables with the intra-hyperfine spin dynamics as reference. In particular, the trapping frequency $\bar{\omega}$ and atom number $N$ only contribute through the mean density $N/V_{\rm{eff}}\propto N^{2/5}\bar{\omega}^{6/5}$, see eqs.~(\ref{eq:Veff}-\ref{eq:RTF}). In this way, the above-described experimental sources of uncertainty in $\omega$ and in $N$ can be combined in a single parameter, which we choose to be the effective trapping frequency $\bar{\omega}_{\rm{eff}}$. In the following calibration, we take $N$ to be the atom number as measured by absorption imaging or Faraday rotation, and obtain $q^{(f)}$ and the effective trap frequency $\bar{\omega}_{\rm eff}$ from a fit to measured $f=1$ intra-hyperfine spin dynamics. This results in a calibration of the QZS and the mean density $N/V_{\rm{eff}}$, now written in terms of measured $N$ and estimated $\bar{\omega}_{\rm{eff}}$. 

\begin{figure}[t!]
\includegraphics[width=1\linewidth]{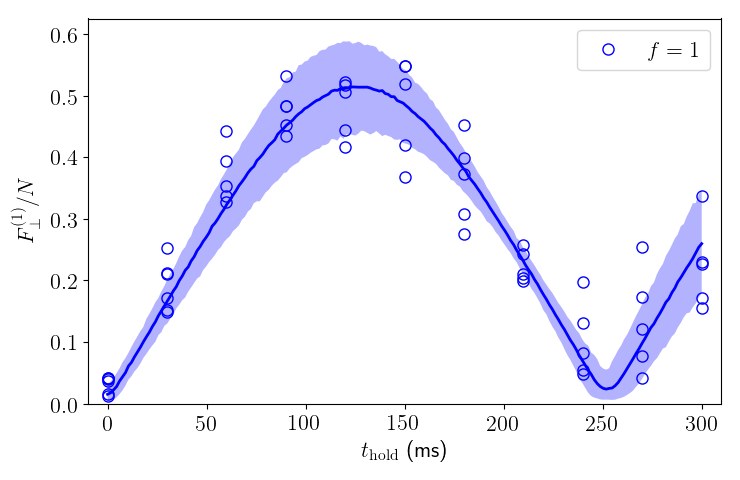} 
\caption{Time evolution of the transverse magnetization in $f=1$, starting from a fully aligned ($F_\perp^{(1)}=0$) state $\xi^{(1)}/\sqrt{N}=(1/2,i/\sqrt{2},1/2)^T$. The dynamics are governed by the competition between the QZS and the ferromagnetic interaction. Blue circles show the observed transverse magnetization $F_{\perp}^{(1)}$ after a variable hold time $t_{\rm hold}$.  The solid line and colored area represent the median and 90\% confidence interval of the theoretical model. The probing is $\delta^{(1)}=\SI{-570}{MHz}$ red-detuned from the $1\leftrightarrow 0'$ transition, such that $G_1^{(1)}=\SI{1.70(12)e-7}{\radian\per spin}$.   } 
\label{fig:trapFreq}
\end{figure}

To this end, we first create a $f=1$ SBEC in the non-magnetic $\xi^{(1)}/\sqrt{N}=(0,1,0)^T$ state, in the presence of a constant field $B=$\SI{119.6}{\milli\gauss}, giving $N =79(4)\times 10^3$ atoms as measured by destructive absorption imaging.  A radio frequency (rf) $\pi/4$ pulse rotates the spin state to $\xi^{(1)}/\sqrt{N}=(1/2,i/\sqrt{2},1/2)^T$.  After a variable hold time, Faraday rotation signals are acquired and fitted with  eqs.~(\ref{eq:faradaySignal}-\ref{eq:faradaySignal2}) to find the transverse magnetization $F_{\perp}^{(f)}$. Results are shown in \cref{fig:trapFreq} and exhibit the expected oscillation of $F_{\perp}^{(f)}$ produced by competition between the QZS and the ferromagnetic interaction. These $F_{\perp}^{(f)}$ values are compared to SMA mean-field simulations as per \cref{sec:NumericalIntegration}, with the $q^{(1)}$ and $\bar{\omega}_{\rm eff}$ as free fit parameters.  We find $\bar{\omega}_{\rm eff}=2\pi\times \SI{90(9)}{\hertz}$ and $q^{(1)}/h= \SI{0.89(10)}{\hertz}$.  

The $\bar{\omega}_{\rm eff}$ value is consistent with independent measurements of trap sloshing frequencies. The obtained value for $q^{(1)}/h$ is in agreement with the theoretically expected $[p^{(1)}/h]^2/\nu_{\rm hfs} = \SI{1.03}{\hertz}$, where $\nu_{\rm hfs}= \SI{6.8}{\giga\hertz}$ is the $f=1,2$ hyperfine splitting. We note that, to the precision of this work, the hyperfine manifolds feature opposite QZS, so that $q^{(2)}/h=\SI{-0.89(10)}{\hertz}$. 

Through \cref{eq:E1} the estimated value of $\bar{\omega}_{\rm eff}$ depends on the ferromagnetic interaction coefficient $g_1^{(1)}$ and thus on $a_{2}^{(1)}-a_{0}^{(1)}$. As mentioned in \cref{sec:errEstimate}, this dependence is undesirable and our preferred quantity to report is the rescaled mean trapping frequency $\bar{\omega}_{\rm eff} |a_{2}^{(1)}-a_{0}^{(1)}|^{5/6}=\SI{1.63(12)E-6}{\second^{-1}\meter^{5/6}}$, which does not depend on the intra-hyperfine interaction.  

\section{Measurement of inter-hyperfine interaction parameters}\label{sec:inter-hyperfine}

We now describe our strategies for measuring $g_1^{(12)}$ and $g_2^{(12)}$, the inter-hyperfine interaction parameters that appear in \cref{eq:E12}.  First, we note that $g_2^{(12)} \overline{P_1^{(12)}}$, which in general is quite complicated, greatly simplifies in the case of stretched spin states in the $f=2$ manifold, for which all $\xi_m^{(2)}$ elements are zero except for either  $\xi_{-2}^{(2)}$ or $\xi_{2}^{(2)}$. For these states $g_2^{(12)} \overline{P_1^{(12)}}$ reduces to a single term, which describes an effective LZS plus an 
effective QZS acting upon the $f=1$ manifold.  As already seen in \cref{fig:trapFreq}, the QZS causes oscillations of $F_{\perp}^{(1)}$. Because the QZS-ferromagnetic competition is the only source of such oscillations, this provides an unambiguous signal by which to measure $g_2^{(12)}$.   

To measure $g_1^{(12)}$, we note that $g_1^{(12)} F_z^{(1)}F_z^{(2)}$ describes an effective LZS of the $f=1$ levels, with a strength proportional to $F_z^{(2)}$, the $f=2$ magnetization along the $\mathbf{B}$-field. The $f=1$ magnetization similarly produces a LZS in the $f=2$ manifold.  The resulting modification of the Larmor frequency is of the order of \SI{1}{\hertz}, a tiny fraction of the $|p^{(f)}/h|\approx \SI{84}{\kilo \hertz}$ LZS due to the external magnetic field $B=\SI{119.6}{\milli\gauss}$. To accurately resolve this shift and decouple the measurements from external magnetic field noise, we operate our SBEC as a comagnetometer. This technique, in which a signal is simultaneously acquired from distinct but co-located sensors, can efficiently reject magnetic field noise while retaining sensitivity to other effects. In hot vapors, comagnetometer techniques have been used for sensing rotation \cite{05Romalis,18Romalis} and searches for physics beyond the standard model \cite{SmiciklasPRL2011}. 

Here, the two sensors are the $f=1$ and $f=2$ manifolds of the SBEC.  Their precession angles $\theta\supf$ have opposite dependences on $\pmin$, and the summed precession angle $\theta\suponetwo \equiv \theta\supone + \theta\suptwo$ is sensitive to $g_1^{(12)}$ with vanishing $\pmin$ contribution. We note that in the SMA the magnetic dipole-dipole interaction produces a field within the condensate that is equally experienced by the $f=1$ and $f=2$ manifolds, and thus with no effect on $\theta\suponetwo$.

\subsection{Interaction parameter $g_2^{(12)}$}
\label{sec:inter-hyperfine1}

To measure $g_2^{(12)}$, we first prepare the state
\begin{equation}\label{eq:initialState1}
\frac{\xi_0}{\sqrt {N}} \equiv \frac{\xi_0^{(1)}\oplus\xi_0^{(2)}}{\sqrt{N}} =
\begin{pmatrix}
\frac{1}{2\sqrt{2}}\\[0.5em]
\frac{i}{2}\\[0.5em]
\frac{1}{2\sqrt{2}}
\end{pmatrix}
\oplus
\begin{pmatrix}
0\\
0\\
0\\
0\\
\frac{1}{\sqrt{2}}
\end{pmatrix}
\; ,
\end{equation}
which describes and equal superposition of $f=1$ in an aligned ($F_\perp^{(1)}=0$) state and $f=2$ in a $-{\bf z}$ stretched state. After a variable wait time the $f=1$ transverse magnetization $F_{\perp}^{(1)}$ is measured by Faraday rotation, as in \cref{sec:calib}. Note that the $f=2$ state is unchanged by the evolution and readout of $f=1$. After measuring the $f=1$ manifold, a rf $\pi /2$ pulse rotates the stretched $f=2$ state into the transverse plane and $F_{\perp}^{(2)}$ is measured by Faraday rotation. This provides a measure of the atom number $N=2 N^{(2)}=101(9)\times 10^3$ atoms. The procedure is described in detail in \cref{sec:appendix1}.

\begin{figure}[t]
\includegraphics[width=1\linewidth]{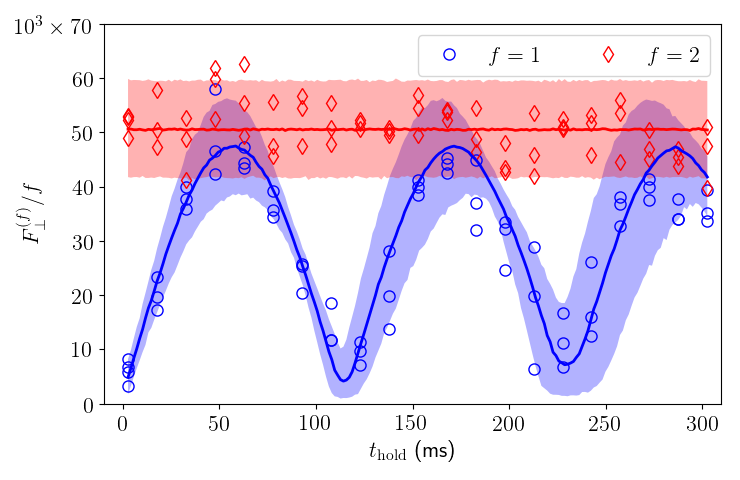} 
\caption{Dynamical evolution for the initial state in \cref{eq:initialState1} and under a magnetic field of \SI{119.6}{\milli\gauss}. Blue circles and red diamonds, are the mean experimental transverse magnetization in $f=1$ and $f=2$, respectively. The solid lines are the median of the theoretical mean-field evolution for $g_2^{(12)}/g_{1}^{(1)}=\SI{-6.4}{}$, while shaded areas represent the 90\% confidence intervals.}
\label{fig:scattLen1}
\end{figure}

In \cref{fig:scattLen1} the measured transverse magnetization in $f=1$ and $f=2$ are shown as a function of the  hold time in the trap. Note that the frequency and amplitude of the modulation nearly double those in \cref{fig:trapFreq}, where only the $f=1$ manifold is populated. By fitting the expected SMA mean-field evolution  we obtain $g_2^{(12)}/g_{1}^{(1)}=\SI{-6.4(6)}{}$.

\subsection{Interaction parameter $g_1^{(12)}$}\label{sec:inter-hyperfine2}

\begin{figure}[t!]
\includegraphics[width=250pt]{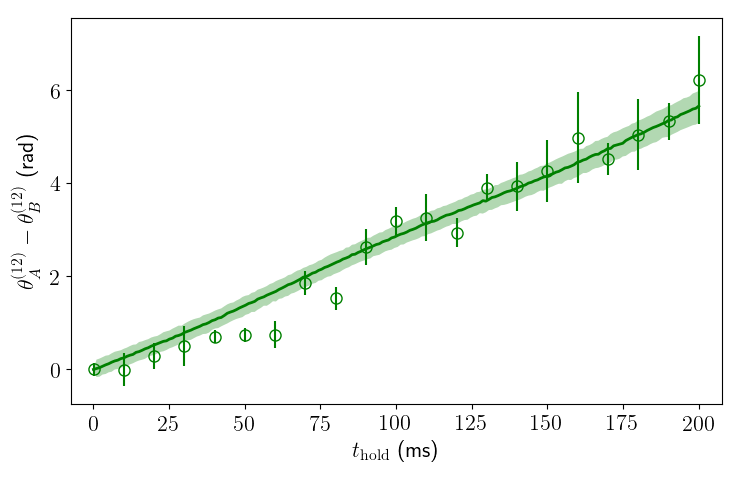} 
\caption{Phase evolution of the difference in the comagnetometer readouts between state preparations A and B. Circles show the experimental mean values and the error bars represent one standard deviation. The solid line and shaded area are the mean and standard deviation of the theoretical phase evolution for $g_2^{(12)}/g_{1}^{(1)}=\SI{-6.4}{}$ and $g_1^{(12)}/g_{1}^{(1)}=\SI{-1.27}{}$. Atom numbers are estimated by destructive absorption imaging before the first run of experimental sequence $A$ and $B$, yielding $N_A=88(3)\times 10^3$ and $N_B=96(3)\times 10^3$. }
\label{fig:scattLen2}
\end{figure}

To measure  $g_1^{(12)}$, we first prepare one of the following two states

\begin{equation}\label{eq:initialState2}
\frac{\xi_{0,A}}{\sqrt{N}}= \hat{R}_{x,\frac{\pi}{6}} \left[
\begin{pmatrix}
\frac{1}{\sqrt{2}}\\
0\\
0
\end{pmatrix}
\oplus
\begin{pmatrix}
\frac{1}{\sqrt{2}}\\
0\\
0\\
0\\
0
\end{pmatrix}\right] , \tag{19a}
\end{equation}

\begin{equation}
\frac{\xi_{0,B}}{\sqrt{N}}= \hat{R}_{x,\frac{\pi}{6}}\left[
\begin{pmatrix}
0\\
0\\
\frac{1}{\sqrt{2}}
\end{pmatrix}
\oplus
\begin{pmatrix}
0\\
0\\
0\\
0\\
\frac{1}{\sqrt{2}}
\end{pmatrix}
\right]
,\tag{19b}
\end{equation}
where $\hat{R}_{x,\frac{\pi}{6}}$ is a rotation about the $\mathbf{x}$ axis by angle $\pi/6$.  The rotation angle  is a compromise between a strong spin component parallel to the external magnetic field (required for a $g_1^{(12)}F_z^{(1)}F_z^{(2)}$ contribution in \cref{eq:E12}) and a significant transverse magnetization (required for Faraday readout).  After a variable wait time, the $f=1$ and $f=2$ precession angles are measured by Faraday rotation.  A detailed description is given in \cref{sec:appendix2}.  

For an initial state $X \in \{ A, B \}$ the comagnetometer signal $\theta_X^{(12)} \equiv \theta_X^{(1)}+\theta_X^{(2)}$ contains contributions from the $-2\poff/h = \SI{-334}{\hertz}$ differential LZS between $f=1$ and $f=2$, the QZS and the spin-dependent inter-hyperfine interaction, i.e., the $g_1^{(12)}$ and $g_2^{(12)}$ contributions. We analyze the difference in comagnetometer readouts $\theta_A^{(12)}-\theta_B^{(12)}$, in which also the differential LZS contribution cancels. The QZS is known from the calibration of \cref{sec:calib}. The results are shown in \cref{fig:scattLen2}, where the experimental data are fitted to SMA mean-field simulations in which $g_1^{(12)}$ is a free fit parameter whereas $g_2^{(12)}$ is fixed at the value found in \cref{sec:inter-hyperfine1}. We obtain $g_1^{(12)}/g_{1}^{(1)}=\SI{-1.27(15)}{}$.

\section{Comparison with prior work and outlook}\label{sec:results}
Using \cref{eq:scattCoefg112} and \cref{eq:scattCoefg212} for the above values of $g_1^{(12)}$ and $g_2^{(12)}$, we find $(a_{3}^{(12)}-a_{2}^{(12)})/(a_{2}^{(1)}-a_{0}^{(1)})=\SI{-1.27(15)}{}$ and $(a_{1}^{(12)}-a_{2}^{(12)})/(a_{2}^{(1)}-a_{0}^{(1)})=\SI{-1.31(13)}{}$, with relative uncertainties of 12\% and 10\%, respectively.  As noted above, these ratios are insensitive to the exact value of $a_{2}^{(1)}-a_{0}^{(1)}$, which serves as an input parameter in the modeling and fits.  The same sensitivity to $a_{2}^{(1)}-a_{0}^{(1)}$ applies also to the prior measurements of \cite{18Hirano}, which found $(a_{3}^{(12)}-a_{2}^{(12)})/(a_{2}^{(1)}-a_{0}^{(1)})=\SI{-1.8(5)}{}$ and $(a_{1}^{(12)}-a_{2}^{(12)})/(a_{2}^{(1)}-a_{0}^{(1)})=\SI{-2.2(4)}{}$. These differ by $1\sigma$ and $2\sigma$ combined uncertainty from the result presented here.   

Our accuracy is presently limited by uncertainty in the SBEC atom numbers, which reflect loading fluctuations and atom loss during the experiment. Active control schemes can stabilize the atom numbers of cold atomic ensembles below shot noise by using dispersive probing \cite{16Sherson}. Applied to the current experiment, such stabilization is foreseen to reduce the relative uncertainties in the results bellow $\SI{\sim 0.3}{\%}$. 

\section{Conclusions}\label{sec:conclusions}
We have demonstrated an interferometric method to precisely measure the inter-hyperfine collisional interactions in $f=1$, $f=2$ mixtures of ultracold atoms. The method employs a single-domain SBEC and hyperfine-state-specific Faraday rotation to measure spin evolution. Two new multi-pulse radio frequency and microwave state preparations are used. Each one generates a hyperfine-state mixture that gives high-visibility spin dynamics, that sensitively depends on one or more inter-hyperfine scattering lengths. We also describe a new calibration of the effective trapping frequency and quadratic Zeeman shifts which is based on the interaction-dependent modulation of the transverse magnetization. This new calibration substitutes for an absolute calibration of the atom number, typically one of the larger uncertainties in ultracold gas experiments.  Applying these techniques to $^{87}$Rb, we measure $(a_{3}^{(12)}-a_{2}^{(12)})/(a_{2}^{(1)}-a_{0}^{(1)})$ and $(a_{1}^{(12)}-a_{2}^{(12)})/(a_{2}^{(1)}-a_{0}^{(1)})$ with relative uncertainties of 12\% and 10\%, respectively, limited by atom number drift between calibration and measurement.  A relative uncertainty of $\SI{\sim 0.3}{\%}$ is projected for experiments with nondestructive atom number monitoring. The methods are directly applicable to other commonly-used alkali species $^{7}$Li, $^{23}$Na and $^{39}$K, in addition to $^{87}$Rb.  

\section{Acknowledgments}
We thank T. Hirano for useful discussions on hyperfine-dependent state preparation. This work was supported by European Research Council (ERC) projects AQUMET (280169), ERIDIAN (713682); European Union projects QUIC (Grant Agreement no. 641122) and FET Innovation Launchpad UVALITH (800901); the Spanish MINECO projects MAQRO (Ref. FIS2015-68039-P), XPLICA (FIS2014-62181-EXP) and QCLOCKS (PCI2018-092973), the Severo Ochoa programme (SEV-2015-0522); Agencia de Gestio d'Ajuts Universitaris i de Recerca (AGAUR) project (2017-SGR-1354); Fundacio Privada Cellex, Fundacio Privada MIR-PUIG and Generalitat de Catalunya (CERCA program). Quantum Technology Flagship project MACQSIMAL (820393); EMPIR project USOQS (17FUN03), Marie Sklodowska-Curie ITN ZULF-NMR (766402).

\bibliographystyle{apsrev4-1no-url}
\bibliography{biblio/biblio}{}

\onecolumngrid
\newpage
\appendix 
\section{Experimental sequences} \label{sec:appendix}

\begin{figure}[h!]
\centering
\includegraphics[width=1.0\linewidth ]{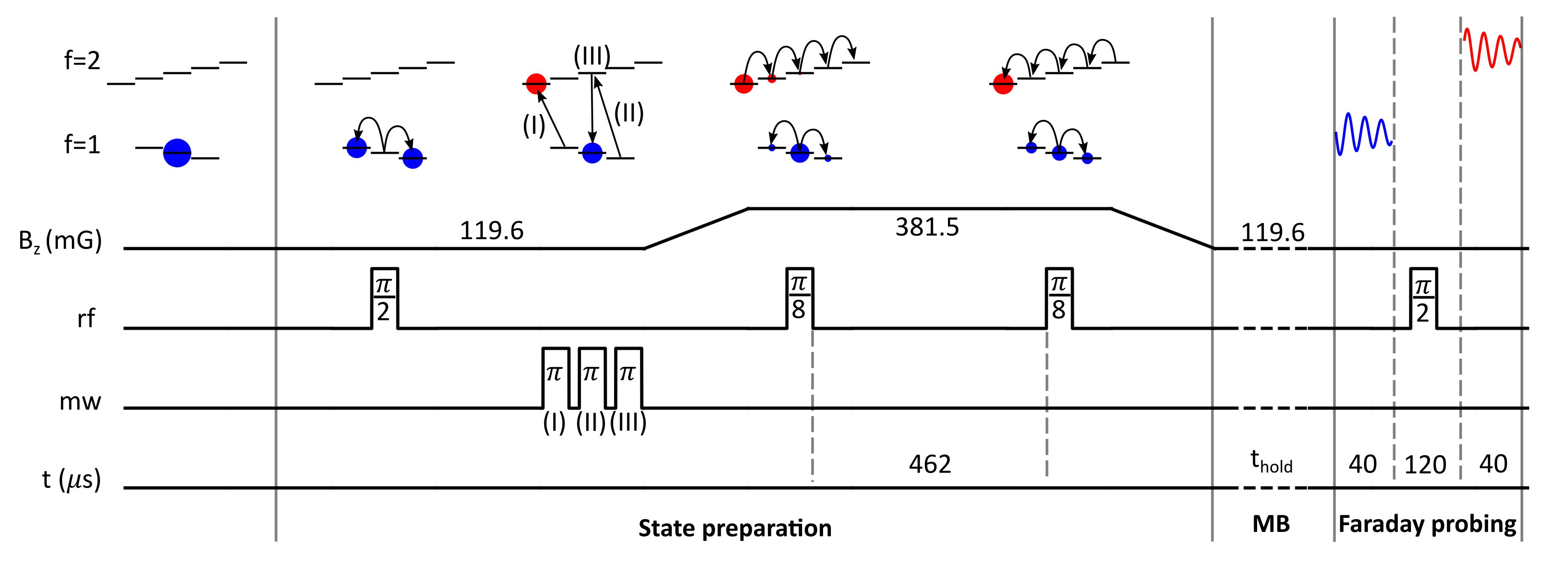} 
\noindent\rule{\textwidth}{1pt}
\includegraphics[width=0.90\linewidth]{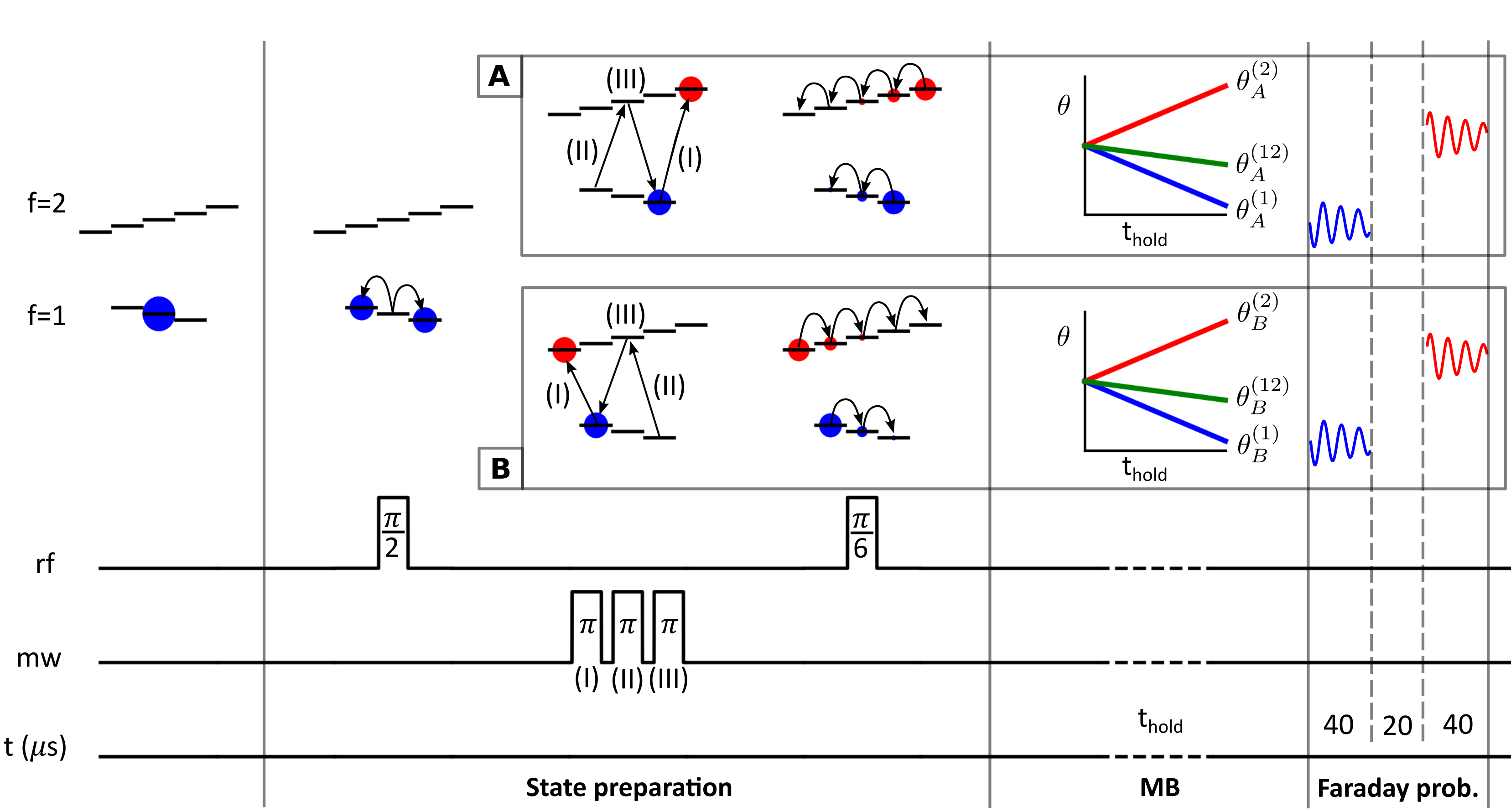} 
\caption{Experimental sequences for the determination of the inter-hyperfine interaction parameters $g_2^{(12)}$ (top) and $g_1^{(12)}$ (bottom). The relative spin populations in $f=1$ and $f=2$ are represented by the blue and red circles. The oscillatory traces at the end of both sequences illustrate the consecutive Faraday readouts of the transverse magnetization in each hyperfine manifold. Note that the bottom sequence branches into A or B for the many-body (MB) evolution of a SBEC initially prepared in $\xi_{0,A}$ or $\xi_{0,B}$, respectively. The insets of this later sequence show the angular evolution of the transverse spins $\theta_A^{(f)}$ and $\theta_B^{(f)}$, as well as the corresponding comagnetometer readouts ($\theta_A^{(12)}$ and $\theta_B^{(12)}$).}
\label{fig:sequence}
\end{figure}

\subsection{Experimental sequence for measuring $g_2^{(12)}$} \label{sec:appendix1}
(See \cref{fig:sequence} top.) After all-optical evaporation, a SBEC is obtained in the $\xi^{(1)}/\sqrt{N}=(0,1,0)^T$ state. The ensemble is coherently transferred into an equal superposition $\xi/\sqrt{N}=(0,1/\sqrt{2},0)^T\oplus(0,0,0,0,1/\sqrt{2})^T$ by means of a resonant radio frequency (rf) $\pi/2$ rotation around the x-axis and a sequence of microwave (mw) $\pi$ pulses (I, II and III). Thereafter, the magnetic field is ramped up to \SI{381.5}{\milli\gauss} in order to raise the differential LZS to $(p^{(1)}+p^{(2)})/h=$\SI{-1.06}{\kilo \hertz}. A Ramsey-like sequence, consisting of two rf $\pi/8$ pulses (rotating about $\mathbf{x}$) separated by \SI{462}{\micro \second} is used to produce a net $\pi/4$ rotation of the $f=1$ manifold, and zero net rotation of the $f=2$ manifold. The resulting state is given in \cref{eq:initialState1}. The magnetic field is rapidly ramped down to \SI{119.6}{\milli\gauss}, ensuring a modest QZS during the subsequent many-body (MB) evolution. After a variable hold time $t_{\rm hold}$, the magnetizations in $f=1$ and $f=2$ are detected by Faraday rotation. A first pulse ($\delta^{(1)}=\SI{-270}{\mega\hertz}$ red detuned from $1\rightarrow 0'$, $G_1^{(1)}=\SI{3.3(2)e-7}{\radian\per spin}$) probes the $f=1$ transverse magnetization. A rf $\pi/2$ pulse is then applied to rotate the  $f=2$ stretched state $\xi^{(2)}/\sqrt{N}=(0,0,0,0,1/\sqrt{2})^T$ into the transverse plane for detection with a second pulse ($\delta^{(2)}=\SI{360}{\mega\hertz}$ blue detuned from $2\rightarrow 3'$, $G_1^{(2)}=\SI{1.9(1)e-7}{\radian\per spin}$). The damped oscillatory signals illustrate the recorded Faraday signals described in eq.~(\ref{eq:faradaySignal}-\ref{eq:faradaySignal2}).

\subsection{Experimental sequence for measuring $g_1^{(12)}$} \label{sec:appendix2}
(See \cref{fig:sequence} bottom.) The sequence starts with a SBEC in $\xi^{(1)}/\sqrt{N}=(0,1,0)^T$ which is coherently split by a rf $\pi/2$ pulse into $\xi^{(1)}/\sqrt{N}=(1/\sqrt{2},0,1/\sqrt{2})^T$. Subsequently, either the initial state $\xi_{0,A}$ or $\xi_{0,B}$ is prepared via mw pulses (I, II and III) and a rf $\pi/6$ rotation around the $\mathbf{x}$ axis. Hereafter, the many-body (MB) evolution begins.  For the applied constant magnetic field of \SI{119.6}{\milli \gauss} the LZS is $p^{(2)}/h\approx -p1^{(1)}/h=\SI{84}{\kilo \hertz}$ with a differential frequency of $p^{(1)}/h+p^{(2)}/h= \SI{-334}{\hertz}$. The insets illustrate how the $f=1$ and $f=2$ transverse spin orientations ($\theta_A^{(1)}$ and $\theta_A^{(2)}$ or $\theta_B^{(1)}$ and $\theta_B^{(2)}$) rapidly evolve due to the LZS. The differential is represented by the green comagnetometer readouts, which depending on the state preparation are labeled by $\theta_A^{(12)}$ and $\theta_B^{(12)}$. After a variable hold time of up to \SI{200}{\milli\second}, the transverse magnetization is interrogated	. First the Faraday probe of $f=1$ is applied, from which, depending on the state preparation, the spin orientation $\theta_A^{(1)}$ or $\theta_B^{(1)}$ is obtained. Next, and without any additional rf pulse, the $f=2$ manifold is probed, yielding $\theta_A^{(2)}$ or $\theta_B^{(2)}$. The comagnetometer readout is obtained by $\theta_X^{(12)} \equiv \theta_X^{(1)}+\theta_X^{(2)}$, where $X \in \{ A, B \}$. Faraday probing frequencies and atom-light coupling factors are identical to the previous section.

\end{document}